\documentclass[aps,11pt,letterpaper,nofootinbib,showpacs,amsmath,amssymb]{revtex4}%
\usepackage{latexsym}%
\usepackage[dvips,dvipdfmx,hypertex]{hyperref}%

\def\tanh{{\text{tanh}}}

\def\cH{\mathcal{H}}
\def\cP{\mathcal{P}}
\def\oP{\overline{P}}
\def\oT{\overline{T}}
\def\oQ{\overline{Q}}
\def\tH{\widetilde{\mathcal{H}}}
\def\tN{\widetilde{N}}
\DeclareMathAlphabet{\mathpzc}{OT1}{pzc}{m}{it}

\newcommand{\beq}{\begin{equation}}
\newcommand{\beqn}{\begin{equation}\nonumber}
\newcommand{\eeq}{\end{equation}}
\newcommand{\bea}{\begin{eqnarray}}
\newcommand{\bean}{\begin{eqnarray}\nonumber}
\newcommand{\eea}{\end{eqnarray}}
\newcommand{\ba}{\begin{align}}
\newcommand{\ea}{\end{align}}

\begin{document}

\title{Canonical Chern-Simons Gravity}
\author{Souvik Sarkar\footnote{\tt sarkarsi@mail.uc.edu}}
\author{Cenalo Vaz\footnote{\tt Cenalo.Vaz@uc.edu}}
\affiliation{Department of Physics, University of Cincinnati, Cincinnati, OH 45221-0011.}

\begin{abstract}
We study the canonical description of the axisymmetric vacuum in 2+1 dimensional gravity, treating Einstein's gravity 
as a Chern Simons gauge theory on a manifold with the restriction that the dreibein is invertible. Our treatment is in 
the spirit of Kucha\v r's description of the Schwarzschild black hole in 3+1 dimensions, where the mass and angular 
momentum are expressed in terms of the canonical variables and a series of canonical transformations are performed
that turn the curvature coordinates and their conjugate momenta into new canonical variables. In their final form, 
the constraints are seen to require that the momenta conjugate to the Killing time and curvature radius vanish and 
what remains are the mass, the angular momentum and their conjugate momenta, which we derive. The Wheeler-DeWitt 
equation is trivial and describes time independent systems with wave functions described only by the total mass and 
total angular momentum.

\pacs{}
\end{abstract}
\maketitle

\section{Introduction}

In 2+1 dimensions, many of the problems associated with quantum gravity are expected to be alleviated by the fact 
that pure gravity in 2+1 dimensions has no local, propagating degrees of freedom. Still, the theory is far from trivial 
\cite{sc1}. The vacuum solutions of pure gravity are multiconical spacetimes, obtained by identification of points in 
flat space \cite{djt} and, in the presence of a cosmological constant, one obtains maximally symmetric solutions, {\it viz.,} 
the Anti-de Sitter (AdS) and de Sitter (dS) spacetimes with a similar identification of points. Such an identification, 
by a discrete subgroup of $SO(2,2)$ in AdS spacetime, was shown to give a spinning black hole solution by Ba\~nados, Teitelboim 
and Zanelli (BTZ) \cite{btz}. The BTZ black hole solution is locally AdS but globally it is characterized by conserved 
charges at the boundary of the AdS spacetime \cite{bhtz}. The solution exhibits many of the properties of black holes 
in 3+1 dimensions and therefore provides a simpler setting for the study of quantum effects.
Likewise, gravitational collapse in 2+1 dimensions is rich in structure. The earliest study of gravitational collapse
in 2+1 dimensions with and without a cosmological constant was carried out in \cite{roma93}. In the context of circularly 
symmetric, homogeneous dust, the authors showed that collapse to a black hole depends sensitively on the initial data. In the
absence of a cosmological constant or in dS spacetime, collapse may or may not occur depending on the initial velocity, 
but if the dust ball collapses then it does so to a naked, conical, point source singularity \cite{gak84}. On the other hand, 
in AdS spacetime the BTZ black hole arises naturally as the end state provided that the initial density is sufficiently large. 
If not, the end state is a again naked conical singularity, but in AdS spacetime. These results led to a numerical study of 
critical phenomena associated with the collapse process in \cite{precho00} and were confirmed in studies of inhomogeneous 
dust collapse in \cite{sg}. Attempts at the quantization of dust collapse in \cite{mw98,vgks} also had several lessons to 
teach. Our ultimate goal is to obtain a description of quantum gravitational collapse in 2+1 dimensions with rotation,
for which a classical description was developed in \cite{vk08}. This paper  is a first step in this program.

Classical 2+1 dimensional gravity and supergravity can also be viewed as Chern-Simons gauge theories of the Poincar\'e, 
Anti-de Sitter and de Sitter groups and their supersymmetric generalizations \cite{ew,kwv}. The general procedure is to identify 
an appropriate (super)group which contains the structure group of the corresponding gravity theory in its even part, 
construct its Lie algebra with generators ${\widehat T}_a$, expand the gauge superfield $A_\mu = A^a_\mu {\widehat T}_a$, 
and construct the Chern-Simons action according to
\bea
I_{C.S.} &=& \frac 12 \text{Tr} \int A \wedge \left( dA + \frac 23 A \wedge A \right)\cr 
&=& \frac 12 \gamma_{ab} \int A^a \wedge \left( dA^b + \frac 13 f^b_{cd} A^c \wedge A^d \right),
\label{cs}
\eea
where $f^a_{bc}$ are the structure constants of the Lie algebra, and $\gamma_{ab} = \text{Tr}(T_a T_b)$ plays the role of 
a metric on the Lie algebra and must be non-degenerate so that the action contains a kinetic term for all components 
of the gauge field.  By construction the action is invariant under a gauge transformation given by
\beq
\delta_g A_\mu = - D_\mu \Lambda,
\eeq
where $\Lambda = \Lambda^a T_a$ and $D_\mu = \partial_\mu + [A_\mu,~~]$ and the classical equations of motion assert 
that the field strengths vanish identically.  For the action \eqref{cs} to be an acceptable gauge theory of (super)gravity, 
gauge transformations must be equivalent to diffeomorphisms. This is indeed true for small diffeomorphisms on-shell. 
It would be incorrect, however, to conclude that Einsein's action in 2+1 dimensions is equivalent to 
the Chern-Simons action because the latter contains many solutions that have no metric interpretation. Here we confine 
our attention to a subspace of solutions that {\it do} have a metric interpretation. The gauge fields are the dreibein 
and the spin connection, and the vanishing field strengths simply assert that the torsion vanishes and the curvature is 
constant. Our aim in this work is to cast the dynamics of Chern-Simons gravity into a canonical form for metric 
compatible solutions, in the spirit of Kucha\v r \cite{kk94}. There is a long history of other approaches in the literature
\cite{other}. These approaches focus on solving the contraints and using them to derive a simplified Hamiltonian in 
a finite number of degrees of freedom, or on exploiting the local isometries to begin with a reduced action for the system. 
The advantage of our approach is that it focuses primarily on simplifying the contraints via a series of canonical 
transformations. These transformations are then easily modified and continue to be useful in simplifying the constraints 
in a variety of systems, including the Einstein-Maxwell system \cite{lwh96}, Lovelock gravity \cite{lswh97}, and when 
matter is included \cite{vws01}. This makes it better adapted to the study of dynamical collapse.

In section II we review the canonical form of the Chern-Simons action for $SO(2,2)$. We employ a general ADM metric to choose 
a natural canonical chart consisting of the three functions comprising the spatial metric, $L(r)$, $R(r)$ and $Q(r)$ and 
their conjugate momenta. We solve three of the six constraints of the Chern-Simons action (corresponding to the vanishing of 
torsion) and show that the other three are equivalent to the Hamiltonian and momentum constraints that would be obtained from 
the second order (Einstein) action. In section III we consider the equations of motion and recover the well-known classical, 
static solutions describing a spinning particle and the BTZ black hole. We then develop the constraints specific 
to the spinning particle and the BTZ black hole. We discuss the appropriate fall off conditions to be imposed on our 
canonical variables in section IV and determine the boundary action. In section V, by embedding the hypersurfaces from which 
our ADM metric is constructed into the spacetimes describing the spinning particle and the BTZ black hole (derived in section III),
we are able to reconstruct the mass and angular momentum in terms of the canonical variables. This allows us to determine a 
new canonical chart in which the constraints are greatly simplified in section VI. Taking into account the boundary action, we 
perform yet another canonical transformation, leading to a desciption in terms of the area radius, the Killing time, the mass and 
angular momentum (and their conjugates) in section VII. The resulting contraints take on a particularly simple form. When they
are imposed as operator constraints on the Wheeler-DeWitt wave functional, the result is as expected: a time independent state, 
which depends only the ADM mass and angular momentum and, once prepared, remains the same on every spacelike hypersurface . 
We summarize our results in the concluding section VIII.

\section{Chern-Simons gravity}

As mentioned in the introduction, vacuum (super)gravity in $2+1$-D can be described as a gauge theory of the gauge 
groups $ISO(2,1)$ (pure gravity), $SO(2,2)$ (AdS), $SO(3,1)$ (dS) and their supersymmetric extensions, with a Chern-Simons 
action ,
\begin{align}
I_{C.S.}=\frac{1}{2}\int_M\gamma_{ab}A^a\wedge(dA^b+\frac{1}{3}{f^b}_{cd}A^c\wedge A^d),
\end{align}
where $A^a$ is the gauge connection, ${f^a}_{bc}$ are the structure constants of the corresponding group $\mathcal{G}$, $\gamma_{ab}$ 
is the metric of the Lie algebra {\it i.e.,} $\gamma_{ab}=2\text{Tr}({\widehat T}_a{\widehat T}_b)$ and the $\widehat T$'s are generators of the Lie algebra. In what 
follows, letters from the beginning of the roman alphabet, $\{a,b,c...\}$, will be used for group indices, the greek alphabet, 
$\{\alpha,\beta...\}$, for spacetime indices and letters from the middle of the roman alphabet, $\{i,j,k...\}$, for spatial indices.
We take the group $\mathcal{G}$ to be the AdS group $SO(2,2)$, with generators $\widehat{P_a}$ and $\widehat{J_a}$ satisfying 
the following commutation relations
\begin{align}
[\widehat{P_a},\widehat{P_b}]=\Lambda\epsilon_{abc}\widehat{J^c},\hspace{.5in}[\widehat{P_a},\widehat{J_b}]=\epsilon_{abc}
\widehat{P^c},\hspace{.5in}[\widehat{J_a}, \widehat{J_b}]=\epsilon_{abc}\widehat{J^c}
\end{align}
where $\Lambda > 0$ is the cosmological constant and the group indices are raised and lowered using the three dimensional Minkowski 
metric. We expand $A_\mu$ in the basis of generators
\begin{align}
 A_\mu={e^a}_\mu\widehat{P_a}+{\omega^a}_\mu\widehat{J_a}
\end{align}
where ${e^a}_\mu$ and ${\omega^a}_\mu$ are the dreibein and the spin connection respectively. There are  two bilinear invariants 
(Casimirs), namely $ \widehat{P}\cdot\widehat{J}+\widehat{J}\cdot\widehat{P}$ and ${\widehat{P}}^2+{\widehat{J}}^2/\Lambda^2$, which
can be used to determine $\gamma_{ab}$ as
\bea
&&\Lambda~ \text{Tr}(\widehat{J_a}\widehat{J_b}) = \text{Tr}(\widehat{P_a}\widehat{P_b}) = \Lambda \eta_{ab}\\
&&\text{Tr}(\widehat{J_a} \widehat{P_b}) = \text{Tr}(\widehat{P_a} \widehat{J_b}) = \eta_{ab}
\eea
respectively. The first is degenerate in the limit as $\Lambda\rightarrow 0$ and would not produce an acceptable Poincar\'e theory in 
that limit. With the second, the Chern-Simons action can be cast in the form
\begin{align}
I_{C.S.}=\frac{1}{2}\eta_{ab}\int_M d^3x~ \epsilon^{\mu\nu\lambda}\left\{{e^a}_\mu\left[\partial_\nu{\omega^b}_\lambda+{\epsilon^b}_{cd}
\left({\omega^c}_\nu{\omega^d}_\lambda+\frac{\Lambda}{3}{e^c}_\nu{e^d}_\lambda\right)\right]+{\omega^a}_\mu\partial_\nu{e^b}_\lambda
\right\}.
\end{align}
As we are primarily interested in the Hamiltonian formulation, it is covenient to separate the time component in the action, and recast 
it in the form
\begin{align}
I_{C.S.}=\frac{1}{2}\eta_{ab}\int_M d^3x~ \epsilon^{ij}\Big\{{e^a}_t&\left[2\partial_i{\omega^b}_j+{\epsilon^b}_{cd}\left({\omega^c}_i
{\omega^d}_j+\Lambda{e^c}_i{e^d}_j\right)\right]\nonumber\\
&+{\omega^a}_t\left[2\partial_i{e^b}_j+2{\epsilon^b}_{cd}{e^c}_i{\omega^d}_j\right]-{e^a}_i\partial_t{\omega^b}_j-{\omega^a}_i
\partial_t{e^b}_j\Big\},
\label{CSaction}
\end{align}
making it evident that the dreibein and the spin connection are canonical conjugates of one another. In this first order form, if
$\{{e^a}_i,{\omega^a}_i\}$ are treated on an equal footing as configuration space variables the canonical momenta do not involve 
time derivatives of the fields and become primary constraints (they are second class). There are then twelve configuration space 
variables, twelve second class constraints and six first class constraints (the theory has no degrees of freedom). One must proceed 
by following Dirac's procedure for constrained systems. 

Here we will follow a different approach. We take the spacetime to be of the form $\mathbb{R} \times \Sigma$ and choose 
${e^a}_i$ for our configuration space variables. The Chern-Simons action \eqref{CSaction} is equivalent to the first order 
Einstein Hilbert action in the dreibein formulation up to a total derivative so, discarding the total time derivative, we find that 
the momentum conjugate to ${e^a}_i$ is
\begin{align}
{\Pi_a}^i =\eta_{ab}\epsilon^{ij}{\omega^b}_j,
\end{align}
where $\epsilon^{ij}$ is two dimensional Levi-Civita tensor. The Hamiltonian density is then
\begin{align}
\mathcal{H}=-\eta_{ab}\Big\{{e^a}_tF^b[\omega]+{\omega^a}_tF^b[e]\Big\},
\label{gaugehamden}
\end{align}
where
\begin{align}
 F_a[e]&=\epsilon_{acd}\eta^{dm}{e^c}_i{\Pi_m}^i\approx0\nonumber\\
 F_a[\omega]&=\partial_i{\Pi_a}^i+\frac{1}{2}\epsilon_{acd}\left(\epsilon_{kl}\eta^{cm}\eta^{dn}{\Pi_m}^k{\Pi_n}^l+\Lambda
 \epsilon^{ij}{e^c}_i{e^d}_j\right)\approx 0 
 \label{gaugeconst}
\end{align}
are the six constraints of the theory. The first three enforce the vanishing of torsion and the second three require 
the curvature to be constant.

Our next task is to rewrite the constraints above in terms of metric functions. We will eventually be interested in 
axisymmetric solutions, so we consider a general isotropic line element in $\Sigma$, with circular coordinates $(r,\phi)$,
\beq
ds^2 = \gamma_{ij} dx^i dx^j = A^2(r) dr^2 + B^2(r) d\phi^2 + C^2(r) dr d\phi,
\eeq
and foliate the three dimensional spacetime with these leaves, which then also become labeled by a time parameter, $t$. The 
resulting ADM metric,
\begin{align}
 ds^2=\overline{N}^2dt^2-A^2(dr+\overline{N}^rdt)^2-B^2(d\phi+\overline{N}^\phi dt)^2-C^2(dr+\overline{N}^rdt)(d\phi+
 \overline{N}^\phi dt)
 \label{ADM}
\end{align}
can be written more conveniently as
\begin{align}
ds^2=N^2dt^2-L^2(dr+N^rdt)^2-R^2\left(d\phi+N^\phi dt+\frac QR dr\right)^2,
\label{modADM}
\end{align}
with the identifications
\beq
L = \sqrt{A^2-\frac{C^4}{4B^2}},~~ R = B,~~ Q = \frac{C^2}{2B},~~ N^r = \overline{N}^r,~~ N^\phi = \overline{N}^\phi + 
\frac{C^2}{2B^2} \overline{N}^r,~~ N = \overline{N}.
\eeq
A driebein which yields the metric in \eqref{modADM} may be given in lower triangular form,
\begin{align}
{e^a}_\mu=
 \begin{pmatrix}
  N&0&0\\
  N^rL&L&0\\
  N^\phi R&Q&R\\
 \end{pmatrix},
\end{align}
in terms of which the non-vanishing constraints become
\begin{align}
 F_0[e]&:=L{\Pi_2}^r-R{\Pi_1}^\phi+Q{\Pi_1}^r\approx0\nonumber\\
 F_1[e]&:=R{\Pi_0}^\phi+Q{\Pi_0}^r\approx0\nonumber\\
 F_2[e]&:=L{\Pi_0}^r+R'\approx0\nonumber\\
 F_0[\omega]&:=\partial_r{\Pi_0}^r + {\Pi_1}^r{\Pi_2}^\phi-{\Pi_2}^r{\Pi_1}^\phi+\Lambda LR\approx0\nonumber\\
 F_1[\omega]&:=\partial_r{\Pi_1}^r+{\Pi_0}^r{\Pi_2}^\phi - {\Pi_2}^r{\Pi_0}^\phi \approx0\nonumber\\
 F_2[\omega]&:=\partial_r{\Pi_2}^r-{\Pi_0}^r{\Pi_1}^\phi+{\Pi_1}^r{\Pi_0}^\phi\approx0.
\end{align}
We may readily solve the first three constraints, which are purely algebraic. From the third we have ${\Pi_0}^r=-R'/L$. 
Inserting this into the second yields ${\Pi_0}^\phi=QR'/LR$ and, from the first equation, ${\Pi_1}^\phi=\frac{L}{R}{\Pi_2}^r-
\frac{Q}{R}{\Pi_1}^r$. With the three momenta obtained, the remaining three non-trivial constraints read
\begin{align}
 F_0[\omega]&:={\Pi_1}^r{\Pi_2}^\phi-{\Pi_2}^r{\Pi_1}^\phi+\Lambda LR-\left(\frac{R'}{L}\right)'\approx0\nonumber\\
 F_1[\omega]&:=\partial_r{\Pi_1}^r-\frac{R'}{L}{\Pi_2}^\phi-\frac{QR'}{LR}{\Pi_2}^r\approx0\nonumber\\
 F_2[\omega]&:=\partial_r{\Pi_2}^r+\frac{R'}{R}{\Pi_2}^r\approx0
\end{align}
and, defining $P_L={\Pi_1}^r$, $P_Q={\Pi_2}^r$ and $P_R={\Pi_2}^\phi$, we may write the simplified Hamiltonian density as 
\beq
\mathcal{H} = -N\mathcal{H}^g - N^r \mathcal{H}_r - N^\phi \mathcal{H}_\phi
\eeq
where 
\bea
&&\mathcal{H}^g = P_LP_R+\Lambda LR-\frac{L}{R}{P_Q}^2+\frac{Q}{R}P_QP_L-\left(\frac{R'}{L}\right)' \approx 0\nonumber\\
&&\mathcal{H}_r = L{P_L}'- R' P_R - \frac{QR'}R P_Q \approx 0\nonumber\\
&&\mathcal{H}_\phi = (RP_Q)' \approx 0,
 \label{const}
\eea
which are the Hamiltonian and momentum constraints of the theory. The last momentum constraint implies that
\begin{align}
R P_Q=\alpha(t),
 \label{phiconst}
\end{align}
and we could also write
\beq
\cH_r =  L{P_L}'- R' P_R + Q P_Q' - \frac QR \cH_\phi \approx 0.
\eeq
To summarize, the phase space is six dimensional, parametrized by three metric functions, $L$, $R$ and $Q$, and their 
conjugate momenta. Axisymmetric solutions are obtained by taking $Q = 0$ and circularly symmetric solutions by taking 
$Q = N^\phi = 0$. The entire content of the theory is in the constraints; the equations of motion follow by taking Poisson 
brackets with $\mathcal{H}$ and, in the following section, we recover the well known stationary solutions with which we 
will work in later sections.

\section{Hamiltonian Equations of Motion}

The Hamiltonian equations of motion are quite generally given by taking Poisson brackets with the smeared constraints,
\bea
&&\dot X = \{X,-H^g[N] - H_r[N^r] - H_\phi[N^\phi]\}_{P.B.}\\
&&\dot P_X = \{P_X,-H^g[N]-H_r[N^r]-H_\phi[N^\phi]\}_{P.B.}.
\eea
For the six phase space variables, 
\begin{align}
 \dot{L}&=\{L,H\}_{P.B.}=-NP_R-\frac{NQ}{R}P_Q+(N^rL)'\nonumber\\
 \dot{R}&=\{R,H\}_{P.B.}=-NP_L+N^rR'\nonumber\\
 \dot{Q}&=\{Q,H\}_{P.B.}= N\left(\frac{2L}{R}P_Q-\frac{Q}{R}P_L\right)+N^r\frac{QR'}R+{N^\phi}'R\nonumber\\
 \dot{P_L}&=\{P_L,H\}_{P.B.}=N\Lambda R-\frac{N'R'}{R^2}-\frac{N{P_Q}^2}{R}+N^r{P_L}'\nonumber\\
 \dot{P_R}&=\{P_R,H\}_{P.B.}=N\Lambda L-\frac{N''}{L}+\frac{N'L'}{L^2}+\frac{NL{P_Q}^2}{R^2}-\frac{Q}{R^2}P_QP_L+(N^rP_R)'-{N^\phi}'
 P_Q\nonumber\\
 \dot{P_Q}&=\{P_Q,H\}_{P.B.}=\frac{N}{R}P_QP_L+N^r{P_Q}'=-\frac{\dot{R}}{R}P_Q
 \label{eoms}
\end{align}
We have made no assumptions about the canonical variables apart from isotropy, so any isotropic, classical solution must satisfy 
\eqref{const} and \eqref{eoms}. Combining the third constraint with the last equation of motion we find that $\alpha$ must be constant.

In the static case the time derivative of all canonical variables must vanish. Using \eqref{phiconst}, this implies that 
the equations of motion, together with the first two constraints \eqref{const}, will read
\begin{align}
 &P_R = \frac{(N^rL)'}{N}-\frac{\alpha Q}{R^2}\nonumber\\
 &P_L = \frac{N^rR'}{N}\nonumber\\
 &N\left(\frac{2\alpha L}{R^2}-\frac{Q}{R}P_L\right)+N^r\frac{QR'}R+{N^\phi}'R = 0\nonumber\\
 &N\Lambda R-\frac{N'R'}{L^2}-\frac{\alpha^2N}{R^3}+N^rP_L' = 0\nonumber\\
 &N\Lambda L-\frac{N''}{L}+\frac{N'L'}{L^2}+\frac{\alpha^2NL}{R^4}-\frac{\alpha Q}{R^3}+(N^rP_R)'-\frac{\alpha}{R}{N^\phi}' = 0\nonumber\\
 &P_LP_R+\Lambda LR-\frac{\alpha^2L}{R^3}+\frac{\alpha Q}{R^2}P_L-\left(\frac{R'}{L}\right)' = 0\nonumber\\
 &{P_L}'-\frac{R'}{L}P_R-\frac{\alpha QR'}{LR^2} = 0
\end{align}
We now have to find eight unknown functions from the above seven equations. So there is the freedom to choose one of the unknown functions 
and we choose $N^r=0$. This gives $P_L=0$, $P_R=-\alpha Q/R^2$ and the last equation is satisfied identically. We are left with four equations 
in five unknowns, namely
\begin{align}
 &\frac{2\alpha NL}{R^2}+{N^\phi}'R=0\nonumber\\
 &N\Lambda R-\frac{N'R'}{L^2}-\frac{\alpha^2N}{R^3}=0\nonumber\\
 &N\Lambda L-\frac{N''}{L}+\frac{N'L'}{L^2}+\frac{\alpha^2NL}{R^4}-\frac{\alpha Q}{R^3}-\frac{\alpha}{R}{N^\phi}'=0\nonumber\\
 &\Lambda LR-\frac{\alpha^2L}{R^3}-\left(\frac{R'}{L}\right)'=0
\end{align}
Therefore we can yet choose another function, this we take to be $Q=0$. Solving the first equation, ${N^\phi}'=-2\alpha NL/R^3$, 
we find that the remaining equations are
\begin{align}
 &N\Lambda R-\frac{N'R'}{L^2}-\frac{\alpha^2N}{R^3}=0\nonumber\\
 &N\Lambda L-\frac{N''}{L}+\frac{N'L'}{L^2}+\frac{3\alpha^2NL}{R^4}=0\nonumber\\
 &\Lambda LR-\frac{\alpha^2L}{R^3}-\left(\frac{R'}{L}\right)'=0
\end{align}
The equations are once again not independent: the second can be obtained from the remaining two equations, so there are 
two independent equations for three unknown functions
\begin{align}
 &N\Lambda R-\frac{N'R'}{L^2}-\frac{\alpha^2N}{R^3}=0\nonumber\\
 &\Lambda LR-\frac{\alpha^2L}{R^3}-\left(\frac{R'}{L}\right)'=0
 \label{finaleom}
\end{align}
and we are free to choose yet another function. We take $R(r) = r$ below.

\subsection{$\Lambda = 0$: The spinning point particle}

With $\Lambda=0$ the gauge group $SO(2,2)$ turns into the Poincar\'e group by a Wigner-Inonu contraction. Taking  $R(r) =r$ the 
equations \eqref{finaleom} readily yield the following solutions
\begin{align}
 L(r)&=\frac{1/\mu}{\sqrt{1+\frac{\alpha^2}{\mu^2r^2}}}\nonumber\\
 N(r)&=N_+\sqrt{1+\frac{\alpha^2}{\mu^2r^2}}\nonumber\\
 N^\phi(r)&=N^\phi_+ +\frac{N_+\alpha}{\mu r^2}
\end{align}
where $\mu$, $N_+$ and $N^\phi_+$ are constants of the integration. For example, if we choose $N_+ = 1$ and $N^\phi_+ =0 $, 
the line element is given by
\begin{align}
 ds^2=N^2 dt^2-\frac{N^{-2}}{\mu^2}dr^2-r^2\left(d\phi-\frac j{ \mu r^2}dt\right)^2
\end{align}
where $\mu$ can be identified with the mass of the particle and $j=-\alpha$ with its angular momentum.

\subsection{$\Lambda\neq 0$: The BTZ Black Hole}

With $\Lambda \neq 0$ we find, from the second of \eqref{finaleom}, that
\begin{align}
 L(r)=\left(\Lambda r^2-M+\frac{\alpha^2}{r^2}\right)^{-1/2}
\end{align}
where $M$ is a constant of the integration. Using this in the first, we have
\begin{align}
 N(r)=N_+\left(\Lambda r^2-M+\frac{\alpha^2}{r^2}\right)^{1/2}
\end{align}
and, together, these imply that $N^\phi=N^\phi_+ + N_+\alpha/r^2$. With $N_+=1$ and $N^\phi_+ = 0$, we recover the BTZ solution 
of mass $M$ and angular momentum $J=-\alpha$ with line element
\begin{align}
 ds^2=N(r)^2dt^2-N(r)^{-2}dr^2-r^2\left(d\phi-\frac{J}{r^2}dt\right)^2
\end{align}

\section{Fall-off conditions and Boundary Action}

The total action in general will combine the bulk action
\beq
S_\Sigma = \int dt \int dr~ \left[P_L \dot L + P_R \dot R + P_Q \dot Q - \cH\right]
\eeq
and a boundary action, $S_{\partial\Sigma}$, whose function is to cancel unwanted boundary variations and whose value will depend 
on the boundary conditions that are imposed. We adopt boundary conditions that enforce every solution to asympotically approach 
one of the spacetimes derived in the previous section. For the maximally extended spinning particle, as for the BTZ black hole, 
the boundaries of spatial hypersurfaces will be taken to lie at $r\rightarrow \infty$.

\subsection{Point Particle}

In case of spinning point particle, we assume that the canonical variables have an asymptotic expansion in integer powers of 
$1/r$ as $r\rightarrow \infty$. We adopt the conditions
\begin{align}
&R \longrightarrow r+\mathcal{O}^{\infty}(r^{-2})\nonumber\\
&L\longrightarrow \frac{1}{\mu_\pm}-\frac{{j_\pm}^2}{2{\mu_\pm}^3}r^{-2}+\mathcal{O}^{\infty}(r^{-3})\nonumber\\
&Q\longrightarrow \mathcal{O}^{\infty}(r^{-2})\nonumber\\
&P_R\longrightarrow P_{R0}+\mathcal{O}^{\infty}(r^{-1})\nonumber\\
&P_L\longrightarrow\mathcal{O}^{\infty}(r^{-1})\nonumber\\
&P_Q\longrightarrow j_\pm r^{-1}+\mathcal{O}^{\infty}(r^{-2})\\
&N\longrightarrow \left[1+\frac{j_\pm}{2\mu_\pm^2}r^{-2}\right] N_\pm + \mathcal{O}^{\infty}(r^{-3})\nonumber\\
&N^r\longrightarrow\mathcal{O}^{\infty}(r^{-1})\nonumber\\
&N^\phi\longrightarrow N^\phi_\pm +\frac{j_\pm}{\mu_\pm}r^{-2}+\mathcal{O}^{\infty}(r^{-3})\nonumber
 \end{align}
where $\mathcal{O}(r^{-n})$ represents a term whose asymptotic behavior is as $r^{-n}$ and is multiplied by some function of $t$
and the plus and minus refer to the right and left infinities respectively. 
It is easily verified that these fall-off conditions are compatible with the constraints and preserved by the time evolution 
equations \eqref{eoms}. To determine the boundary action, we must consider all terms in the Hamiltonian density, $\cH$, whose 
variation will lead to boundary terms. As no derivative of $P_R$ appears, a variation of $P_R$ will yield no boundary contribution 
and, due to the fall-off conditions, contributions from all the variations with respect to $R$ and $P_L$ will fall off much faster 
than $r^{-1}$. Only variations with respect to $L$ and $P_Q$ yield boundary contributions, {\it viz.,} 
\begin{align}
 \int dt \left[NR'\delta\left(\frac{1}{L}\right)-N^\phi R\delta P_Q \right]\Bigg\vert_{\partial M} = -\int dt
 [N_+\delta \mu_+ + N_-\delta \mu_- - N^\phi_+ \delta j_+ - N^\phi_- \delta j_-]
\end{align}
This must be cancelled by an appropriate boundary action, which we therefore take to be
\begin{align}
S_{\partial \Sigma} = \int dt[N_+ \mu_+ + N_-\mu_- - N^\phi_+ j_+ - N^\phi_- j_-]
\label{boundact1}
\end{align}
The boundary action affirms the role of $\mu$ and $-\alpha =j$ as the mass and the angular momentum of the spinning 
particle.

\subsection{BTZ Black Hole}

In this case, for the asymptotic behaviour of our canonical variables we adopt
\begin{align}
 &R\longrightarrow r+\mathcal{O}^\infty(r^{-2})\nonumber\\
 &L\longrightarrow \frac{r^{-1}}{\sqrt{\Lambda}}+\frac{M_\pm r^{-3}}{2\Lambda^{3/2}}+\mathcal{O}^{\infty}(r^{-4})\nonumber\\
 &Q\longrightarrow\mathcal{O}^\infty(r^{-6})\nonumber\\
 &P_L\longrightarrow \mathcal{O}^\infty(r^{-2})\nonumber\\
 &P_R\longrightarrow \mathcal{O}^\infty(r^{-4})\nonumber\\
 &P_Q\longrightarrow -J_\pm r^{-1}+\mathcal{O}^\infty(r^{-2})\nonumber\\
 &N\longrightarrow \left(\sqrt{\Lambda}r-\frac{M_\pm}{2\sqrt{\Lambda}}r^{-1}\right)N_+ +\mathcal{O}^\infty(r^{-2})\nonumber\\
 &N^r\longrightarrow \mathcal{O}^\infty(r^{-2})\nonumber\\
 &N^\phi\longrightarrow N^\phi_\pm + J_\pm r^{-2}+\mathcal{O}^\infty(r^{-4})
 \label{falloffbtz}
\end{align}
Again, it is easy to check that these conditions are compatible with the constraints and preserved by the time evolution equations.
As before, only those variables whose space derivatives appear in the action 
will contribute to the boundary action. From the action we see that $R$, $L$, $P_L$, $P_Q$ are all likely to contribute
to the boundary action. However, by explicitly performing the variation we find that the variation with respect to $R$ and $P_L$ 
rapidly approach zero at both boundaries, but variations with respect to $L$ and $P_Q$ contribute at $r\rightarrow \infty$. Explicitly,
using the asymptotic expressions for corresponding variables as before, we find that the boundary action will be 
\begin{align}
S_{\partial \Sigma} = -\int dt \left[\frac 12 (N_+ M_+ + N_- M_-) + N^\phi_+ J_+ + N^\phi_- J_-\right]
\label{boundact2}
\end{align}
While the inclusion of a boundary action un-freezes the evolution at infinity, it leads to another problem, which is that the lapse 
and shift functions may also be varied at the boundaries. This would lead to the conclusion that $\mu_\pm = j_\pm = M_\pm = J_\pm =0$. 
Therefore Kucha\v r \cite{kk94} proposed that $N_\pm$ and $N^\phi_\pm$ should be viewed as prescribed functions of $t$. This 
``parametrization at infinity'' will be exploited in section VII to present a greatly reduced form of the canonical action. 

\section{Embedding}

Our aim now is to develop the action and constraints specific to the two solutions obtained in section III. First we 
show how the canonical data determine the mass and the angular momentum of these systems by embedding the hypersurfaces 
of the ADM metric into the metrics describing the spinning particle and the BTZ black hole respectively, imagining that 
they are leaves of a particular foliation of these spacetimes.

\subsection{Spinning Particle}

We begin with the spinning point particle, expressing the metric in terms of the Killing time and area radius as
\begin{align}
 ds^2=F dT^2-\frac{1}{\mu^2 F} dR^2-R^2\left(d\phi - \frac{j}{\mu R^2}dT\right)^2
 \label{spinpart}
\end{align}
where $F=\left(1+\frac{j^2}{\mu^2R^2}\right)$, $\mu$ and $j$ are the mass and angular momentum respectively. It is convenient 
rescale the Killing time according to $\oT = T/\mu$; the metric in \eqref{spinpart} can then be written as
\beqn
 ds^2=F_1 d\oT^2-\frac{1}{F_1} dR^2-R^2\left(d\phi - \frac{j}{R^2}d\oT\right)^2
\eeq
where $F_1=\left(\mu^2+\frac{j^2}{R^2}\right)$. Taking $\oT$ and $R$ to be functions of the ADM coordinates, {\it i.e.,} 
$\oT=\oT(t,r)$ and $R=R(t,r)$, and comparing with the ADM form of the line element, $\eqref{modADM}$, we find
\begin{align}
 N&=\frac{R'\dot{\oT}-\oT'\dot{R}}{L}\nonumber\\
 N^r&=\frac{F_1^{-1}\dot{R}R'-F_1\dot{\oT}\oT'}{L^2}\nonumber\\
 N^\phi&=-\frac{j\dot{\oT}}{R^2}\nonumber\\
 L^2&=F_1^{-1}{R'}^2 - F_1{\oT'}^2\nonumber\\
 Q&=-\frac{j\oT'}{R}
 \label{multipliers}
\end{align}
Inserting the lapse and shift into the second equation of \eqref{eoms} we then have
\begin{align}
 P_L&=\frac{1}{N}(-\dot{R}+N^rR')=-\frac{F_1\oT'}{L}~~ \Rightarrow~~ T'=-\frac{LP_L}{F_1}
\end{align}
which, inserted into the expression for $L^2$ in \eqref{multipliers}, gives
\begin{align}
 F_1 &= \mu^2 + \frac{j^2}{R^2} = \left(\frac{{R'}^2}{L^2}-P_L^2\right)
 \label{F1}
 \end{align}
and, from the last equation in \eqref{multipliers}, we also find,
\begin{align}
 j &=\frac{QR}{LP_L}\left(\frac{{R'}^2}{L^2}-P_L^2\right).
\end{align}
Together, these equations allow us to recover the mass and angular momentum from the canonical data. Furthermore, differentiating 
\eqref{F1} with respect to $r$, we find
\begin{align}
 (F_1)'=\left(\frac{R'^2}{L^2}-P_L^2\right)'&=-2P_LP_L'+2\left(\frac{R'}{L}\right)\left(\frac{R'}{L}\right)'\nonumber\\
 &=-\frac{2R'}{L}\cH_g-\frac{2P_L}L\cH_r-\frac{2R'}{R}P_Q^2 
\end{align}
where $\cH^g$ and $\cH_r$ are the Hamiltonian and momentum contraints. Therefore,
\beq
\left(\mu^2+\frac{j^2}{R^2} - P_Q^2\right)' =  -\frac{2R'}{L}\cH_g-\frac{2P_L}L\cH_r - \frac{2P_Q}R\cH_\phi
\label{keyeqn1}
\eeq
is a linear combination of the constraints, which, we note, do not require $\mu'$ and $j'$ to separately vanish.

\subsection{BTZ Black Hole}

Similarly, for the the BTZ black hole, the metric is expressed as 
\begin{align}
F(R) dT^2-\frac{1}{F(R)}dR^2-R^2\left(d\phi-\frac{J}{R^2}dT\right)^2
\end{align}
where $F(R) =\Lambda R^2-M+\frac{J^2}{R^2}$, $M$ and $J$ are the mass and angular momentum respectively. Embedding $\eqref{modADM}$ into 
BTZ metric we obtain \eqref{multipliers}. Then inserting the lapse and shift into the second equation of \eqref{eoms} we obtain $T'$
and substitute its value into the expressions for $L^2$ and $Q$; this gives
\begin{align}
 F&= \Lambda R^2 - M + \frac{J^2}{R^2} = \frac{{R'}^2}{L^2}-{P_L}^2\nonumber\\
 J&=\frac{QR}{LP_L}\left(\frac{{R'}^2}{L^2}-{P_L}^2\right)
 \label{FandJ}
\end{align}
and, again, one recovers the mass and angular momentum in terms of the canonical data. Furthermore,
\begin{align}
(F-\Lambda R^2)' =\left(\frac{R'^2}{L^2}-P_L^2-\Lambda R^2\right)' &= -2P_LP_L'-2\Lambda RR'+2\left(\frac{R'}{L}\right)
\left(\frac{R'}{L}\right)'\nonumber\\
 &=-\frac{2R'}{L}\mathcal{H}^g-\frac{2P_L}L\mathcal{H}_r-\frac{2R'}{R}P_Q^2
\end{align}
and we may write
\beq
\left(-M + \frac{J^2}{R^2}-P_Q^2\right)' = -\frac{2R'}{L}\mathcal{H}^g-\frac{2P_L}L\mathcal{H}_r 
- \frac{2P_Q}R \cH_\phi \approx 0.
\label{keyeqn2}
\eeq
As before, the constraints do not require $M'$ and $J'$ to separately vanish. Equations \eqref{keyeqn1}
and \eqref{keyeqn2} are identical apart from the sign of the mass terms. If these solutions are regarded 
as end states of collapse then the signs strongly depend on the initial data, as discovered in \cite{roma93}
and noted in the introduction.

\section{New Canonical Variables}

We have determined the mass and angular momentum in terms of the canonical variables. Following Kucha\v r \cite{kk94}, we now seek a 
new set of canonical variables in which the constraints are simplified. From the expressions for $\mu~ (M)$ and $j~ (J)$, 
however, it appears that both the mass and the angular momentum cannot be a part of the same canonical chart because their 
Poisson brackets do not vanish. In the quantum theory, they are not simultaneously observable. A more transparent configuration
space variable is provided by the time-time component of the metrics. We will show how this comes about.

We will work with a non-zero cosmological constant as the spinning particle is the $\Lambda\rightarrow 0$ limit of the same
together with $M\rightarrow -\mu^2$. From the expression for $F$ in \eqref{FandJ} it is straightforward to show that 
\beq
Z = \frac{R'^2}{L^2} - P_L^2- \Lambda R^2 - P_Q^2,~~ P_Z=-\frac{LP_L}{2F}
\eeq
are conjugate variables, {\it i.e.,} $\{Z,P_Z\}_{P.B.}=1$. However the Poisson brackets of $Z$ with $Q$ and $P_R$,
as well as the Poisson bracket of $P_Z$ with $P_R$ are non-vanishing, 
\beq
\begin{matrix}
&\{Z,R\}_{P.B.}=0 &\hskip 0.75in &\{Z,P_R\}_{P.B.}=-\left(\frac{2R'}{L^2}\right)'-2\Lambda R\cr
&\{Z,Q\}_{P.B.}=2P_Q &\hskip 0.75in &\{Z,P_Q\}_{P.B.}=0\cr
&\{P_Z,R\}_{P.B.}=0 &\hskip 0.75in &\{P_Z,P_R\}_{P.B.}=-\left(\frac{R'P_L}{F^2L}\right)'\cr
&\{P_Z,Q\}_{P.B.}=0 &\hskip 0.75in &\{P_Z,P_Q\}_{P.B.}=0\end{matrix}
\eeq
We wish to replace $L$ and $P_L$ by $Z$ and $P_Z$ in the canonical chart and the Poisson brackets above tell us that 
a canonical transformation to new variables, $\oP_R$ and $\oQ$ is required. However, one explicitly checks that
\beq
\oQ = Q + \frac{L P_L P_Q}{F}
\label{oQ}
\eeq
does, in fact, have vanishing Poisson brackets with $Z,P_Z$ and $R$ and is conjugate to $P_Q$. The only remaining problem 
is to find $\oP_R$ and there is a standard procedure for achieving this. The canonical transformation from the original chart, 
$\{L,R,Q,P_L,P_R,P_Q\}$, to the new chart, $\{Z,R,\oQ,P_Z,\oP_R,P_Q\}$, is found to be generated by
\beq
G[L,R,P_L,P_Q] = \int dr \left[LP_L \left(1- \frac{P_Q^2}{F}\right) - R' \tanh^{-1}\left(\frac{R'}{LP_L}\right)\right],
\eeq
and $\oP_R$ is determined to be
\beq
\oP_R = P_R - \frac{\Lambda R L P_L}F - \frac{(R'/LP_L)'}{1-(R'/LP_L)^2}.
\eeq
The fall-off of the new canonical variables is easily determined from the fall-off conditions \eqref{falloffbtz}.

The momentum constraint $\cH_r$, written in terms of the new variables, now reads,
\beq
\cH_r = Z'P_Z-R'\oP_R+\oQ P_Q' -\left(\frac{\oQ-2P_ZP_Q}R\right) \cH_\phi
\eeq
and, by substituting the new variables into the Hamiltonian constraint, we also find
\beq
\cH^g = \frac{2FP_Z}{RL}\left[\oQ P_Q + R \oP_R\right] -\frac{R'}{2FRL}\left[2P_Q (R P_Q)'+RZ'\right].
\eeq
This last expression can be greatly simplified by exploiting \eqref{keyeqn1}; after some algebra we find
\beq
\cH^g = +\frac{F}{RP_L}\left[\oQ P_Q + R \oP_R\right] - \frac{R'}{L^2P_L} \cH_r,
\eeq
so the full Hamiltonian can now be written in terms of new constraints,
\begin{align}
 \tH^g&= R\oP_R + \oQ P_Q \nonumber\\
 \tH_r&=Z'P_Z - R'\overline{P}_R + \oQ P_Q' .\nonumber\\
 \cH_\phi &= (R P_Q)' .
 \label{constraintsf}
\end{align}
and adjoined to the action by means of new multipliers. Explicitly,
\beq
\cH = -\tN \tH^g - \tN^r \tH_r - \tN^\phi \tH_\phi
\eeq
where
\bea
\tN &=& \frac{NF}{RP_L}\cr\cr
\tN^r &=& N^r + \frac{NR'}{L^2P_L}\cr\cr
\tN^\phi &=& N^\phi - \left(N^r+\frac{NR'}{L^2P_L}\right)\left(\frac\oQ R\right)
\label{lapseshiftf}
\eea
with $\oQ$ given in \eqref{oQ}. We also notice that 
\beq
\frac{R'}R \tH^g + \tH_r - \frac\oQ R \cH_\phi = Z'P_Z,
\eeq
so we could just as well consider the constraint system
\begin{align}
 \tH^g&= R\oP_R + \oQ P_Q \nonumber\\
 \cH_Z&=Z'P_Z\nonumber\\
 \cH_\phi &= (R P_Q)' 
 \label{constraintsf2}
\end{align}
and adjoin these (instead of \eqref{constraintsf}) to the bulk action by means of new multipliers. In the next section we will 
absorb the boundary action into the bulk action and by doing so we will be able to simplify the constraint system even further.

\section{The Boundary Action}

We will make one more canonical transformation, a trivial one interchanging coordinates and momenta,
\beq
\oQ = -P_Y,~~ P_Q = Y.
\eeq
In terms of the new variables, the Chern Simons action takes the form
\beq
S[Z,R,\Omega;P_Z,\oP_R,P_\Omega] = \int dt \int_{-\infty}^\infty dr\left[P_Z\dot Z  + \oP_R\dot R  + P_Y\dot Y  - \tN\tH^g - 
\tN^r \tH_r - \tN^\phi\tH_\phi \right] + S_{\partial\Sigma}
\eeq
where the boundary action is given by \eqref{boundact2} in terms of the old variables. If the lapse and shift functions on the 
boundary were allowed to be freely varied, it would imply that the mass and angular momentum of the black hole both vanish 
at infinity. To avoid this conclusion and allow for a non-vanishing mass and angular momentum, they must be treated as 
prescribed functions of the ADM time parameter, $t$, {\it i.e.,} the lapse and shifts must have fixed ends. To determine these
functions, we compare the asymptotic ADM metric in \eqref{modADM} at fixed $r$,
\beq
ds^2 = (N_\pm^2-R^2 {N^\phi_\pm}^2) dt^2 - 2R^2 N^\phi_\pm dtd\phi - R^2 d\phi^2
\eeq
with the asymptotic metric in comoving coordinates \cite{vk08} (also at fixed $r$)
\beq
ds^2 = d{\mathfrak t}_\pm^2 + 2 \Omega_\pm R^2 d\tau d\phi -R^2 d\phi^2,
\eeq
where ${\mathfrak t}$ is the proper time and $\Omega$ is the angular velocity. Evidently, we must take
\beq
N_\pm = \pm \sqrt{1+r^2\Omega_\pm^2}~ \dot {\mathfrak t}_\pm~ \stackrel{\text{def}}{=}~ \pm\dot\tau_\pm ,~~ N^\phi_\pm = \mp 
\Omega_\pm \dot{\mathfrak t}_\pm~ \stackrel{\text{def}}{=}~ \pm\dot\omega_\pm
\eeq
where ${\mathfrak{t}}_\pm(t)$ represents the proper time and $\Omega_\pm(t)$ the angular velocity function as measured along 
constant $r$ world lines at the infinities. The surface action now reads
\beq
S_{\partial\Sigma} = -\int dt\left[\frac 12\left(M_+\dot\tau_+ - M_-\dot\tau_-\right) + J_+\dot \omega_+ - J_-\dot \omega_-\right].
\eeq
First, consider the Liouville form 
\beq
\Theta_1 := \int_{-\infty}^\infty dr~ P_Z\delta Z  - \frac 12\left(M_+\delta\tau_+ - M_-\delta\tau_-\right)
\eeq
and note that, according to the fall-off conditions, $\lim_{r\rightarrow \infty} Z(r) = -M_\pm$. We therefore define the 
function $\Gamma(r)$ by
\beq
Z(r) = -M_- - \int_{-\infty}^r dr'~ \Gamma(r'),~~ Z'(r) = -\Gamma(r)
\eeq
and rewrite $\Theta_1$ as follows:
\bea
\Theta_1 &:=& \int_{-\infty}^\infty dr P_Z(r)\left[-\delta M_- - \int_{-\infty}^r dr' \delta\Gamma(r')\right] - \frac 12 
\delta (\tau_+ M_+ - \tau_- M_-) + \frac 12\tau_+\delta M_+  - \frac 12\tau_-\delta M_- \cr\cr
&=&\delta M_- \left(-\frac 12 \tau_- - \int_{-\infty}^\infty dr P_Z(r)\right) - \int_{-\infty}^\infty dr P_Z(r)\int_{-\infty}^r dr' 
\delta\Gamma(r')\cr
&& \hskip 2cm +\frac 12 \tau_+ \left(\delta M_- + \int_{-\infty}^\infty dr' \delta\Gamma(r')\right) - \frac 12 \delta (\tau_+
M_+ - \tau_- M_-)\cr
&=& \delta M_- \left(\frac 12 (\tau_+ - \tau_-) - \int_{-\infty}^\infty dr P_Z(r)\right) - \int_{-\infty}^\infty dr P_Z(r)\int_{-\infty}^r dr' 
\delta\Gamma(r')\cr
&& \hskip 2cm +\frac 12 \tau_+ \int_{-\infty}^\infty dr' \delta\Gamma(r') - \frac 12 \delta (\tau_+ M_+ - \tau_- M_-)
\eea
This allows us to identify the conjugate variables, 
\beq
m = M_-,~~ p_m = \frac 12 (\tau_+ - \tau_-) - \int_{-\infty}^\infty dr P_Z(r)
\eeq
and we have, apart from an exact form,
\beq
\Theta_1 = p_m\delta m + \int_{-\infty}^\infty dr \left[\frac 12 \tau_+ \delta \Gamma(r) - P_Z(r)\int_{-\infty}^r dr' \delta\Gamma(r')\right].
\eeq
Again, using the identity \cite{kk94},
\beq
\int_{-\infty}^\infty dr P_Z(r)\int_{-\infty}^r dr' \delta\Gamma(r') = - \int_{-\infty}^\infty dr~ \delta \Gamma(r)\int_{\infty}^r dr' 
P_Z(r'),
\eeq
we find
\beq
\Theta_1 = p_m\delta m + \int_{-\infty}^\infty dr \left[\frac 12 \tau_+ + \int_{\infty}^r dr' P_Z(r')\right]\delta \Gamma(r)
\eeq
which now allows us to identify the conjugate variables
\beq
\Gamma(r) = -Z'(r),~~ P_\Gamma(r) = \frac 12 \tau_+ + \int_{\infty}^r dr' P_Z(r')
\eeq
We note that $P_\Gamma' =  P_Z = \frac 12 T'$, so the Killing time can be identified with the momentum $2P_\Gamma$ up to a constant.
We can choose the constant so that $T$ matches $\tau_+$ at infinity, then 
\beq
T = 2 P_\Gamma = \tau_+ + 2\int_{\infty}^r dr' P_Z(r')
\eeq
and the momentum conjugate to the Killing time is just $\cP_T = -\frac 12\Gamma$.

Next, consider the Liouville form
\beq
\Theta_2 := \int_{-\infty}^\infty dr~\left[\oP_R \delta R + P_Q\delta \oQ\right] - \left(J_+\delta\omega_+ - J_-\delta\omega_-\right)
\eeq
and recall that, under the fall-off conditions, $\lim_{r\rightarrow\infty} P_Q = -J_\pm r^{-1}$. If we define
\beq
R(r)P_Q(r) = -J_- + \int_{-\infty}^r dr' \Sigma(r')
\eeq
then, by the third constraint, $\Sigma(r)=0$. Therefore $J_+=J_- = J$ and 
\bea
\Theta_2 &:=& \int_{-\infty}^\infty dr~\left[\oP_R \delta R - \frac JR\delta \oQ\right] - J\delta\left(\omega_+ - \omega_-\right)\cr
&=& \int_{-\infty}^\infty dr~\left(\oP_R+\frac{J\oQ}{R^2}\right) \delta R - J\delta\left[\left(\omega_+ - \omega_-\right)+
\int_{-\infty}^\infty dr~\left(\frac\oQ R\right)\right]\cr
&=&\int_{-\infty}^\infty dr~\left(\oP_R+\frac{J\oQ}{R^2}\right) \delta R + \left[\left(\omega_+ - \omega_-\right)+
\int_{-\infty}^\infty dr~\left(\frac\oQ R\right)\right]\delta J
\eea
up to an exact form. Thus we identify
\beq
p_J = \left(\omega_+ - \omega_-\right)+\int_{-\infty}^\infty dr~\left(\frac\oQ R\right)
\eeq
as the momentum conjugate to $J$ and a new momentum,
\beq
\cP_R = \oP_R + \frac{J\oQ}{R^2},
\eeq
conjugate to $R$. In terms of the new variables, the constraints read
\begin{align}
 \cH_R&= R\cP_R \nonumber\\
 \cH_T&=T'\cP_T
 \label{constraintsT}
\end{align}
and may be adjoined to the canonical action by means of new Lagrange multipliers. The reduced canonical action,
\beq
S = p_m \dot m + p_J \dot J +\int_{-\infty}^\infty dr \left[\cP_T \dot T + \cP_R \dot R  - \left(N^T \cP_T+ N^R \cP_R \right)\right],
\label{reducedaction}
\eeq
shows that the configuration space of vacuum 2+1 dimensional gravity is covered by the coordinates $T$, $R$ and 
two degrees of freedom, $m$ and $J$ . The constraints are starightforward: $\cP_T=\cP_R = 0$. 

Quantization proceeds directly. According to Dirac's quantization program, the momenta are raised to operator status 
and the constraints act as operator constraints on the state functional, $\Psi = \Psi(m,J,t;T,R]$. The two constraints 
tell us that the wave functionals are independent of $T$ and $R$ and the spacetimes are described by wave functions, 
$\Psi(m,J,t)$, which, moreover, are time independent because the Hamiltonian vanishes,
\beq
i \dot \Psi(m,J,t) = 0~ \Rightarrow~ \Psi(m,J,t) = \Psi(m,J).
\eeq
No further information is available. The wave function, once prepared, stays the same on every spacelike hypersurface.

\section{Conclusion}

As mentioned in the introduction, the principal goal of this paper was to construct the canonical description of 
axisymmetric, vacuum solutions to Einstein's gravity in 2+1 dimensions using techniques that are easily extended to 
the description of dynamical collapse. Just as the original canonical reduction of static spherical 
geometries in 3+1 dimensions by Kucha\v r \cite{kk94} has proved extremely useful in understanding spherically symmetric 
dynamical collapse, we expect the reduction presented in this paper to play a pivotal role in describing gravitational 
collapse with rotation, at least in 2+1 dimensions \cite{vk08}.

Here, we showed that the mass and angular momentum can be recovered from the canonical data and that the constraints 
describing the axisymmetric vacuum turn out to be extremely simple after a series of canonical transformations and after 
absorbing the boundary terms into the hypersurface action. Indeed, one finally arrives at a trivial system (with zero 
Hamiltonian) described by the only two classical features of this vacuum: the ADM mass and the angular momentum. The 
quantum mechanics of the system describes a time independent wave function that depends only on these variables.  
This may seem a bit surprising considering that particle production is expected to occur near the horizon and, in fact, 
this state of affairs will no longer hold once matter is injected into the spacetime. Studies of spherical quantum dust 
collapse, employing Kucha\v r's variables, have shown how Hawking radiation arises in this approach \cite{kmhv07}, but 
they have also shown that (modulo a selection rule) the collapse process need not lead to the formation of black holes
\cite{svw16}. Infalling 
shells of matter are accompanied by expanding shells emanating from the center and neither the infalling shells nor the 
expanding shells may ever cross the horizon. 

Our ultimate goal is to couple gravity to matter and describe the quantum 
evolution of gravitational collapse with rotation. One advantage of the canonical approach is that all information is 
preserved in the canonical data. Thus, if the leaves of the foliation are chosen carefully, so that they cover all of the 
spacetime, then one can in principal study the collapse everywhere, without and within the horizon and even in the 
approach to the singularity. Such a foliation is provided by slices of constant proper time and we will report on the 
classical and quantum results from embedding the ADM metric here into the spacetime described in \cite{vk08} in a future 
publication.

\end{document}